\documentclass{article}
\usepackage{spconf,amsmath,graphicx}
\usepackage{amsthm,amssymb}

\usepackage{relsize}
\usepackage[scaled=0.9]{helvet}

\usepackage{color,graphicx}
\usepackage[usenames,dvipsnames]{xcolor}
\usepackage[bookmarks,colorlinks,
			pdfauthor={Guido Carlo Ferrante},%
			pdftitle={TranIT}]{hyperref}%
\definecolor{CiteColor}{rgb}{0.99,0.22,.00}
\definecolor{LinkColor}{rgb}{0.22,0.11,.99}
						\hypersetup{colorlinks,%
						citecolor= red,%
						filecolor= blue,%
						linkcolor= blue,%
						urlcolor= blue}

\usepackage{enumerate}

\def\clap#1{\hbox to 0pt{\hss#1\hss}}

\usepackage{etoolbox}

\theoremstyle{plain}
\newtheorem{theorem}{Theorem} 

\newtheorem{corollary}{Corollary}
\newtheorem{lemma}{Lemma}

\renewcommand{\qedsymbol}{}

\AtEndEnvironment{definition}{\null\hfill\ensuremath{\blacksquare}}%
\AtEndEnvironment{assumption}{\null\hfill\qedsymbol}%
\AtEndEnvironment{rem}{\null\hfill\qedsymbol}%
\AtEndEnvironment{example}{\null\hfill\qedsymbol}%

\usepackage{braket}
\usepackage{booktabs}
\let\geq\geqslant

\usepackage{caption}
\usepackage{eufrak} 
\newcommand{\ubar}[1]{\text{{\kern-0.5ex}\b{$\kern0.5ex{#1}$}}}

\newcommand{\barprec}[2]{\text{{\kern-#1}{$\bar{\kern#1{#2}}$}}}
\IfFileExists{mtpro2.sty}{%
	\IfFileExists{umt2hrb.fd}{%
		\usepackage[noamssymbols,slantedGreek,zswash,mtpscr]{mtpro2}
		}{%
		\usepackage[lite,noamssymbols,slantedGreek,zswash]{mtpro2}
		\usepackage{mathrsfs}
		}%
}{%
\usepackage{mathrsfs}
}%
\usepackage{bm} 
\newcommand{\bs}[1]{\bm{#1}}

\usepackage[bb=boondox]{mathalfa}
\usepackage{rotating,ragged2e}

\usepackage{enumerate}

\newcommand{\numberset}[1]{\text{\usefont{U}{BOONDOX-ds}{m}{n}{#1}}}
\newcommand{\C}{\numberset{C}\hspace{0.1ex}}

\newcommand{\CNorm}[2]{\ensuremath\text{\usefont{OMS}{cmsy}{m}{n}{CN}}(#1,#2)}

\def\nslim{\hspace{-0.2ex}}

\usepackage{mathtools}

\newcommand{\msf}[1]{\mathsf{#1}}

\DeclareMathOperator{\Expectation}{\text{\usefont{U}{BOONDOX-ds}{m}{n}{E}}}

\newcommand{\Einline}[1]{\Expectation[\,#1\,]}
\newcommand{\Espec}[2]{\Expectation{#1[}\,#2\,{#1]}}

\DeclareMathOperator{\diag}{diag}

\DeclareMathOperator{\range}{range}

\def\h{\dagger}

\def\ph{{{\color{white}\h}}}
\def\phn{{{\color{white}\h}\!}}
\def\t{{\msf{T}\!}}
\def\SNR{\msf{SNR}}
\def\ie{\textit{i.e.}}

\def\iexp{\kern0.2ex{\text{\usefont{T1}{cmss}{m}{it}j}\kern0.2ex}}

\usepackage{array}

\let\Gamma\varGamma
\let\Delta\varDelta
\let\Theta\varTheta
\let\Lambda\varLambda
\let\Xi\varXi
\let\Pi\varPi
\let\Sigma\varSigma
\let\Upsilon\varUpsilon
\let\Phi\varPhi
\let\Psi\varPsi
\let\Omega\varOmega

\usepackage{xargs}
\newcommandx{\yaHelper}[2][1=\empty]{%
\ifthenelse{\equal{#1}{\empty}}%
  {\ensuremath{\scriptstyle{#2}}}
  {\raisebox{#1}[0pt][0pt]{\ensuremath{\scriptstyle{#2}}}}
}
\newcommandx{\yrightarrow}[4][1=\empty, 2=\empty, 4=\empty, usedefault=@]{%
  \ifthenelse{\equal{#2}{\empty}}%
  {\xrightarrow{\protect{\yaHelper[#4]{#3}}}}
  {\xrightarrow[\protect{\yaHelper[#2]{#1}}]{\protect{\yaHelper[#4]{#3}}}}
}

\def\toasinline{\,{\yrightarrow{\msf{a.s.}}[-0.4ex]}\,}
\def\toas{\;{\yrightarrow{\ \msf{a.s.}\ }[-0.4ex]}\;}

\newcolumntype{L}[1]{>{\raggedright\let\newline\\\arraybackslash\hspace{0pt}}m{#1}}
\newcolumntype{C}[1]{>{\centering\let\newline\\\arraybackslash\hspace{0pt}}m{#1}}
\newcolumntype{R}[1]{>{\raggedleft\let\newline\\\arraybackslash\hspace{0pt}}m{#1}}

\title{Group-blind detection with very large antenna arrays \\in the presence of pilot contamination}
\name{G.~C.~Ferrante$^{\circ\ast}$, G.~Geraci$^{\circ}$, T.~Q.~S.~Quek$^{\circ}$, and M.~Z.~Win$^{\ast}$}
\address{$^{\circ}$SUTD, Singapore, and $^{\ast}$MIT, MA}
\begin{document}
%
\maketitle
\begin{abstract}
Massive MIMO is, in general, severely affected by pilot contamination. As opposed to traditional detectors, we propose a group-blind detector that takes into account the presence of pilot contamination. While sticking to the traditional structure of the training phase, where orthogonal pilot sequences are reused, we use the excess antennas at each base station to partially remove interference during the uplink data transmission phase. We analytically derive the asymptotic SINR achievable with group-blind detection, and confirm our findings by simulations. We show, in particular, that in an interference-limited scenario with one dominant interfering cell, the SINR can be doubled compared to non-group-blind detection. 
\end{abstract}
%
%


\section{Introduction}

Massive MIMO is, along with network densification and increased bandwidth, one of the key technologies promising to tremendously increase the rate per area in future cellular networks \cite{BocHeaLozMarPop:2014,AndBuzWanetal:2014}. The idea behind massive MIMO is to equip base stations (BSs) with a number of antennas much larger than the number of active users per time-frequency resource block \cite{Lueatl:2014,RusPerBuoLar:2013}. This allows to increase the uplink SNR through maximal-ratio combining, and make the matched-filter detector optimal with perfect channel state information. 

However, %
the modern cellular network architecture foresees the use of orthogonal pilots to estimate the channel between each user and the BS. The number of orthogonal pilots is upper bounded by the coherence time \cite{HasHoc:2003, JinLoz:2010}, hence pilots are usually reused in different cells. As a consequence, channel estimation is corrupted by the interference received during the training phase from users in other cells, a phenomenon known as \textit{pilot contamination} \cite{Mar:2010,NgoLarMar:2013,Josetal:2011}. As the number of antennas at each BS grows, the rate achievable by traditional receivers saturates due to pilot contamination \cite{HoytenDeb:2013,GopJin:2011,FerAshMar:2013,KriYatMan:2012}. %
%
In order to mitigate pilot contamination through an improved channel estimation, a nonlinear iterative algorithm that jointly estimates channels and transmitted symbols has been proposed in \cite{NgoLar:2012}. 
A step toward the understanding of the fundamental limits of massive MIMO has been recently made by M\"uller {et al.} \cite{MulCotVeh:2014,MulVehCot:2013,CotMulVeh:2013}, showing that pilot contamination can be removed if the power received from users within the cell is larger than that received from out-of-cell users. However, this assumption requires both power control and a regular cell geometry, and may not hold in a dense cellular network \cite{AndClaDohetal:2012}. 

In this paper, we propose a receiver design for the uplink of massive MIMO multiuser cellular networks. The proposed receiver takes into account pilot contamination, by adapting the group-blind detection scheme originally proposed in \cite{WanHos:1999} to the case of imperfect channel knowledge due to contamination. Unlike \cite{MulCotVeh:2014,MulVehCot:2013,CotMulVeh:2013}, we do not assume neither power control nor regular cell tessellation. Moreover, this paper differs from \cite{NgoLar:2012} as it focuses on improving data detection rather than modifying the channel estimation phase. 
We provide an asymptotic analysis of the SINR achievable by our scheme, showing a gain with respect to non-group-blind receivers. In particular, we show an SINR gain equal to two in an interference-limited scenario with one dominant interfering cell. Simulations validate our analysis and confirm the rate improvement attained by our scheme.

\section{System Model}
\subsection{Received signal}
Consider the uplink of a noncooperative multicellular network with $L$ cells. Each cell is equipped with one BS having $n$ antennas, each BS serving $K$ single-antenna users on the same time-frequency signaling resource. Throughout the paper, the reference cell is referred to as cell $1$, and interfering cells are labelled with indices $l\in\{2,\dotsc,L\}$. Users in the reference cell and in other cells will be referred to as \textit{in-cell users} and \textit{out-of-cell users}, respectively. The signal received by the reference BS during symbol period $m$ is:
\begin{equation}\label{eq:basic} \bs{y}(m)=\sum_{l=1}^L\sum_{k=1}^K \bs{h}_{lk} \sqrt{\beta_{lk}} \, x_{lk}(m)+\bs{n}(m), \end{equation}
where: $\bs{h}_{lk}=[h_{lk1},\dotsc,h_{lkn}]^\t\in\C^{n}$ is the channel vector between user $k$ in cell $l$ and the reference BS, being $h_{lkr}$ the channel coefficient with respect to antenna $r$; $\beta_{lk}>0$ is the channel gain between the reference BS and user $k$ in cell $l$, that models pathloss and shadowing effects; $x_{lk}(m)$ is the symbol transmitted by user $k$ in cell $l$; $\bs{n}(m)\in\C^{n}$ is additive white Gaussian noise (AWGN) vector. We assume $\{\beta_{kl}\}$ fixed during the coherence time, $h_{lkr}\sim\CNorm{0}{1}$, $\Einline{\bs{n}(m)\bs{n}(m')^\h} = \bs{I}\delta_{mm'}$, $\Einline{\bs{h}_{lk}\bs{h}_{l'k'}^\h} = \bs{I}\delta_{ll'}\delta_{kk'}$, $\Einline{x_{lk}x_{l'k'}} =P\delta_{ll'}\delta_{kk'}$, 
where $P$ is the transmitted power, assumed equal for all users. Denote $\bs{H}_l=[\bs{h}_{l1},\dotsc,\bs{h}_{lK}]\in\C^{n\times K}$, $\bs{R}_l=\diag(\beta_{l1},\dotsc,\beta_{lK})$, and $\bs{G}_l=\bs{H}_l\bs{R}_l^{1/2}=[\bs{g}_{l1},\dotsc,\bs{g}_{lK}]\in\C^{n\times K}$. Compactly, eq.~\eqref{eq:basic} can be written as follows for the generic symbol period $m$:
\begin{align} \bs{y} =\sum_{l=1}^L \bs{H}_{l} \bs{R}_{l}^{1/2} \bs{x}_{l}+\bs{n}=\sum_{l=1}^L \bs{G}_{l} \bs{x}_{l}+\bs{n}, \label{eq:basic2}\end{align}
where dependence on $m$ is implicit.

\subsection{Channel estimation}
By using orthogonal pilots during the training phase, the MMSE estimation $\bs{\hat{g}}_{1k}$ of $\bs{g}_{1k}$ is \cite{Kay:1993,HoytenDeb:2013}
\begin{equation}
\bs{\hat{g}}_{1k} = \bigg(\sum_{l\geq 1} \bs{g}_{lk}+\sqrt{\epsilon}\bs{\nu}_{1k}\bigg)\varphi_{1k}^\phn\beta_{1k}^{-1},
\end{equation}
where $1/\epsilon$ is equal to the effective training SNR, $\bs{\nu}_{1k}\thicksim\CNorm{\bs{0}}{\bs{I}}$, and
 \begin{equation}
\varphi_{1k}=\frac{\beta_{1k}^2}{\epsilon+\sum_{l\geqslant 1}\beta_{lk}}.\label{eq:epsfirst}
\end{equation}
We collect in matrix form estimations $\bs{\hat{G}}_l=[\bs{\hat{g}}_{l1},\dotsc,\bs{\hat{g}}_{lK}]$ and estimation errors $\bs{\tilde{G}}_l=\bs{G}_l-\bs{\hat{G}}_l$. 

\subsection{Achievable rate}
Following \cite{HasHoc:2003,HoytenDeb:2013}, an achievable rate $R_{1k}$ for in-cell user $k$ can be derived by considering the signal
\begin{align} 
\bs{y}' = \bs{\hat{G}}_{1} \bs{x}_{1} + \bs{\tilde{G}}_{1} \bs{\tilde{x}}_{1} + \sum_{l>1} \bs{G}_{l} \bs{x}_{l}+\bs{n}, \label{eq:basic3bis}
\end{align}
where $\bs{\tilde{x}}_{1}$ is independent on $\bs{x}_1$ and has same covariance. Let $\bs{w}_{1k}$ denote the linear receiver for user $k$, the rate $R_{1k}$ is given by 
\begin{equation}\label{eq:achrate}
R_{1k} = \Einline{ \log(1+\gamma_{1k}) },
\end{equation}
where the expectation is with respect to estimated channels, and the SINR $\gamma_{1k}$ is given in \eqref{eq:gamma} at the top of the next page.
\begin{figure*}
\begin{equation}
\gamma_{1k} = \frac{ |\bs{w}_{1k}^\h \bs{\hat{g}}_{1k}^\ph|^2 }{ \Espec{\bigg}{ \displaystyle{\bs{w}_{1k}^\h \bigg( \frac{1}{P}\bs{I} + \bs{\tilde{g}}_{1k}^\ph\bs{\tilde{g}}_{1k}^\h + \sum_{j\neq k} \bs{g}_{1j}^\ph\bs{g}_{1j}^\h + \sum_{l>1}\sum_{j\geqslant 1} \bs{g}_{lj}^\ph\bs{g}_{lj}^\h  \bigg)\, \bs{w}_{1k} \;\bigg|\;\bs{\hat{G}}_{1}} } }.\label{eq:gamma}
\end{equation}
\hrule
\end{figure*} 

\section{Proposed Group-Blind Detector}

Blind receivers were developed for multiuser detection and equalization \cite{Mad:1998}, and then generalized to group-blind detection in the context of CDMA \cite{WanHos:1999,ZhaWan:2002,HosWan:2002,XuWan:2004}. While in blind techniques the receiver knows the signature sequence of the user to decode only, in group-blind techniques it knows the signature sequences of a subset of users. In the uplink of a cellular network, this corresponds to a BS knowing in-cell channels and being unaware of out-of-cell channels. While group-blind detection was originally proposed assuming perfect knowledge of a subset of channels, we hereby extend group-blind detection to the case of contaminated knowledge.

The proposed receiver $\bs{w}_{1k}$ consists of two components. A first component, $\bs{\dot{w}}_{1k}$, belongs to $\range{\bs{\hat{G}}_1}$ and is derived 
on the basis of the signal 
$\bs{y}^{\textup{in}} = \bs{\hat{G}}_{1} \bs{x}_{1} +\bs{n}$; 
the MMSE criterion yields
\begin{equation}
\bs{\dot{w}}_{1k} = (\bs{\hat{G}}_1^\phn\bs{\hat{G}}_1^\h+\textstyle{\frac{1}{P}} \bs{I})^{-1} \bs{\hat{g}}_{1k}.\label{eq:innerrx}
\end{equation}
A second component, $\bs{\breve{w}}_{1k}$, belongs to a subspace orthogonal to $\range{\bs{\hat{G}}_1}$ and that lies within the signal space. Let $\bs{\breve{U}}_{\nslim\bs{\hat{G}}_1}$ be a matrix whose columns span such subspace. The component $\bs{\breve{w}}_{1k}$ is derived by taking into account the whole received signal. Following the MMSE criterion, a derivation similar to \cite{WanHos:1999} that also accounts for imperfect channel estimation due to pilot contamination yields
\begin{equation}
\bs{\breve{w}}_{1k} = -\,\bs{\breve{U}}_{\nslim\bs{\hat{G}}_1}^\ph \! \Big(\bs{\breve{U}}_{\nslim\bs{\hat{G}}_1}^\h\bs{C}_{\bs{y}'}^\phn \bs{\breve{U}}_{\nslim\bs{\hat{G}}_1}^\phn\Big)^{\!{-1}} \bs{\breve{U}}_{\nslim\bs{\hat{G}}_1}^\h \bs{C}_{\bs{y}'}^\ph \bs{\dot{w}}_{1k}^\phn,\label{eq:mywang}
\end{equation}
where $\bs{C}_{\bs{y}'}$ is the covariance matrix of \eqref{eq:basic3bis}. 
The group-blind detector $\bs{w}_{1k} = \bs{\dot{w}}_{1k} + \bs{\breve{w}}_{1k}$ is, therefore, explicitly given by
\begin{equation}
\bs{w}_{1k} = \bigg\{ \bs{I}-\,\bs{\breve{U}}_{\nslim\bs{\hat{G}}_1}^\ph \! \Big(\bs{\breve{U}}_{\nslim\bs{\hat{G}}_1}^\h\bs{C}_{\bs{y}'}^\phn \bs{\breve{U}}_{\nslim\bs{\hat{G}}_1}^\phn\Big)^{\!{-1}} \bs{\breve{U}}_{\nslim\bs{\hat{G}}_1}^\h \bs{C}_{\bs{y}'}^\ph \bigg\} \bs{\dot{w}}_{1k}^\phn.\label{eq:hypoteticalrx}
\end{equation}
We can show that simple blanking techniques \cite{AndBuzWanetal:2014} allow to accurately estimate $\bs{C}_{\bs{y}'}$ for the purpose of implementing \eqref{eq:hypoteticalrx}. Details are omitted due to space constraints.

\section{Asymptotic Performance Analysis}

We derive the asymptotic achievable rate, as $n\to\infty$. In this limit, the SINR is bounded by non-vanishing interference terms. 
In order to obtain these asymptotically non-vanishing terms, we use the two following properties \cite{Mar:2010,HoytenDeb:2013}:
\begin{enumerate}[(i)]
\item channels are asymptotically orthogonal in the almost sure sense, \ie, $n^{-1}\bs{g}_{kl}^\h\bs{g}_{k'l'}\phn\toasinline \beta_{kl}\delta_{kk'}\delta_{ll'}$, where $\delta_{ij}$ denotes the Kronecker delta;
\item in the high-SNR regime, $\bs{\hat{g}}_{1k}\in\mathscr{S}_k=\range\{\bs{g}_{lk}\colon l\geqslant 1\}$.
\end{enumerate}
\begin{figure}[t]\centering
\includegraphics{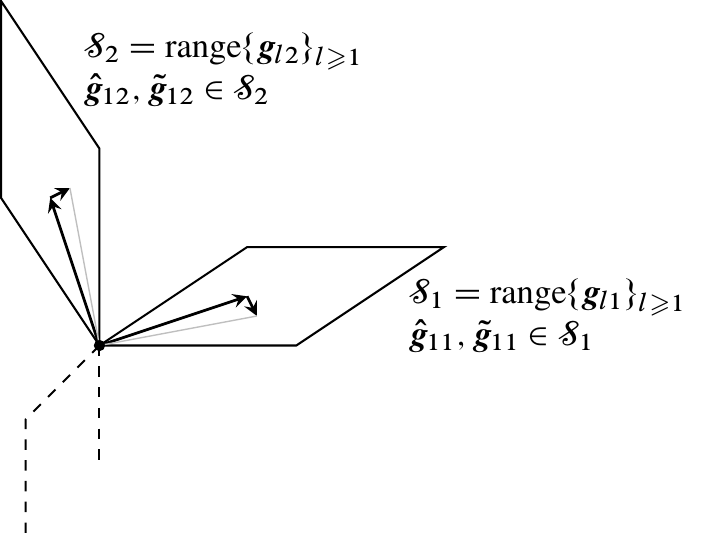}
\caption{Conceptual representation of the asymptotic structure of the signal space in the high-SNR regime. All channels are almost surely orthogonal. For any fixed in-cell user $k$, both estimated and error vectors belong to $\mathscr{S}_k=\range{\!\{\bs{g}_{lk}\colon {l\geqslant 1}\}}$.}
\label{fig:vspace}
\end{figure}
A conceptual representation of the structure of the signal space in the high-SNR regime as $n\to\infty$ is shown on Fig.~\ref{fig:vspace}. Properties (i) and (ii) imply that, asymptotically, the signal space $\mathscr{S}=\range{[\bs{G}_1\cdots\bs{G}_L]}$ is the direct sum $\mathscr{S}_1\oplus\cdots\oplus\mathscr{S}_K$. 
However, the projection of $\bs{\hat{g}}_{1k}$ onto vectors in $\bs{S}_k$ does not asymptotically vanish, in general. In fact, the following limit holds: 
\begin{multline}
\frac{1}{n}\bs{\hat{g}}_{1k}^\h \bs{g}_{lj}^\ph = \frac{1}{n}\varphi_{1k}\beta_{1k}^{-1}\bigg(\sum_{m\geqslant 1} \bs{g}_{mk}^{\nslim\h}+\sqrt{\epsilon}\bs{\nu}_{1k}^\h\bigg) \bs{g}_{lj}\\ \toas \varphi_{1k}^\phn\beta_{1k}^{-1} \beta_{lj} \delta_{kj}.\label{eq:contam}
\end{multline}
%
As a consequence, in-cell user $k$ is asymptotically interfered only by out-of-cell users who used the same training sequence. On the basis of the above observations, we can derive the following results, that hold for the case of $L=2$ cells. We note that such a scenario approximates an interference-limited network with one dominant interfering cell. Proofs as well as analytical results for $L\geq 2$ cells are omitted and will be provided in a journal version of this manuscript. Numerical results that validate our analysis are given in \S~\ref{sec:simres}. 

\begin{lemma}\label{lem:L2} Let $L=2$. Asymptotically, the variable after detection satisfies
\begin{multline}
\hspace{-2ex} \frac{1}{n}\bs{w}_{1k}^\h\bs{y}' 
	\toasinline \varphi_{1k} x_{1k} + \bigg\{(\varphi_{1k}\beta_{1k}^{-1}\beta_{2k})-\frac{(\varphi_{1k}\beta_{1k}^{-1}\beta_{2k})^3}{\lambda}\bigg\} x_{2k} \\
	 + \lambda^{-1} (\varphi_{1k}\beta_{1k}^{-1}\beta_{2k})^2(\beta_{1k}-\varphi_{1k}) \tilde{x}_{1k} \label{eq:lem2}
\end{multline}
where 
$\lambda=(\beta_{1k}-\varphi_{1k})^2+(\varphi_{1k}\beta_{1k}^{-1}\beta_{2k})^2$. 
\end{lemma}

The SINR achieved with group-blind detection readily follows from Lemma~\ref{lem:L2}.

\begin{theorem}\label{th:L2} Let $L=2$. The SINR $\gamma_{1k}$ achieved with the proposed group-blind detector satisfies:
\begin{equation}
\boxed{\gamma_{1k}\toas\bar{\gamma}_{1k}=\bigg[1+\frac{1}{(1+\epsilon/\beta_{2k})^2}\bigg]\rho_{1k}^2} \label{eq:gammaL2}
\end{equation}
where $\rho_{1k}=\beta_{1k}^\phn\beta_{2k}^{-1}$.
\end{theorem}

Let $\bar{\gamma}_{1k}'$ denote the SINR achieved with traditional (non-group-blind) detection, \ie, when $\bs{\breve{w}}_{1k}=\bs{0}$, given by \cite{Mar:2010,HoytenDeb:2013}
\begin{equation}
\bar{\gamma}_{1k}'=\frac{\beta_{1k}^2}{\sum_{l>1}\beta_{lk}^2}.\label{eq:sinrMF}
\end{equation}
We define the asymptotic SINR gain $\bar{\eta}_{1k}$ provided by the proposed group-blind detector as
\begin{equation}
\bar{\eta}_{1k}=\frac{\bar{\gamma}_{1k}}{\bar{\gamma}_{1k}'}.\label{eq:ratio}
\end{equation}
For $L=2$, \eqref{eq:sinrMF} reduces to $\bar{\gamma}_{1k}'=\beta_{1k}^2\beta_{2k}^{-2}$, which combinied with \eqref{eq:gammaL2} and \eqref{eq:ratio} yields
\begin{equation}
\bar{\eta}_{1k}=1+\frac{1}{(1+\epsilon/\beta_{2k})^2}.
\end{equation}
Both $\bar{\gamma}_{1k}$ and $\bar{\eta}_{1k}$ simplify in the limit $\epsilon\to 0$, as specified in the following Corollary. 

\medskip
\begin{corollary}\label{cor:L2} Let $L=2$. The SINR gain obtained by using the group-blind receiver for user $k$ satisfies
\begin{equation}
\bar{\eta}_{1k}\to 2  \ \ \textup{as}\ \ \epsilon\to0,\label{eq:ratioapprox}
\end{equation}
hence the asymtptotic SINR $\gamma_{1k}$ reduces to
\begin{equation}
\boxed{\bar{\gamma}_{1k}\to 2 \bar{\gamma}_{1k}' \ \ \textup{as}\ \ \epsilon\to0.}
\label{eq:newgammaapprox}
\end{equation}
\end{corollary}

The above Corollary shows that, in the high-SNR regime, the asymptotic SINR achieved with group-blind detection is doubled compared to traditional detection.

Let $\Delta \bar{R}_{1k}$ be the difference between the asymptotic rates achieved by user $k$ with and without group-blind detection, respectively, given by
\begin{equation}
\Delta \bar{R}_{1k}= \log(1+\bar{\gamma}_{1k})-\log(1+\bar{\gamma}_{1k}').\label{eq:DeltaR}
\end{equation} 
In the high-SNR regime, $\Delta\bar{R}_{1k}\approx\rho_{1k}^2$ when $\beta_{2k}\gg \beta_{1k}$ (strong out-of-cell interference), while $\Delta \bar{R}_{1k}\approx1$~b/s/Hz when $\beta_{2k}\ll \beta_{1k}$ (weak out-of-cell interference). Note that the case $\beta_{2k}\gg \beta_{1k}$ can occasionally occur when BSs are randomly deployed, resulting in irregular Voronoi cells.

\section{Numerical Results}\label{sec:simres}

%
\begin{figure}[t]
\centering
\includegraphics{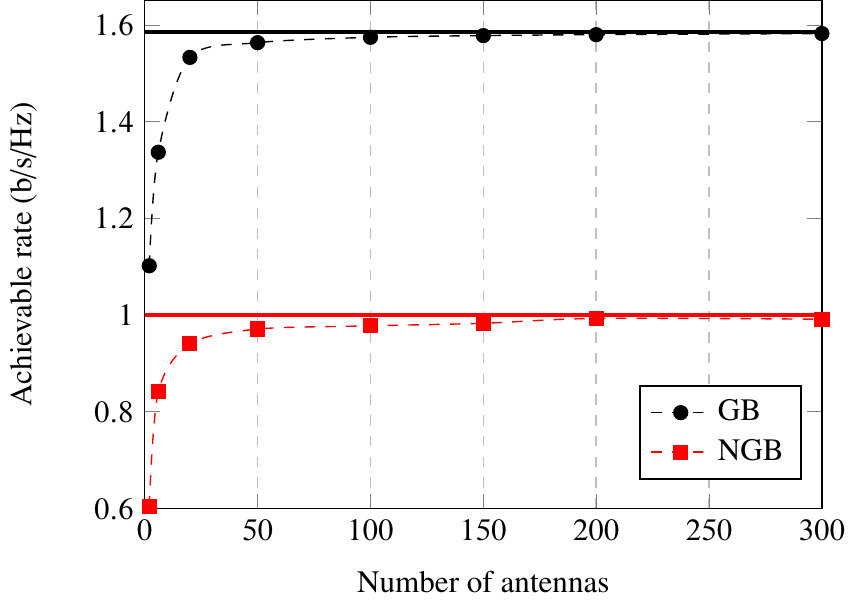}
\caption{Achievable rate (b/s/Hz) as a function of the number of antennas $n$ with and without group-blind detection. Scenario parameters: $L=2$, $K=1$, $\SNR=20$ dB and $\beta_{11}/\beta_{21}=0$~dB (strong interference).}
\label{fig:fig3}
\end{figure}
\begin{figure}[t]
\centering
\vspace{-6px}
\includegraphics{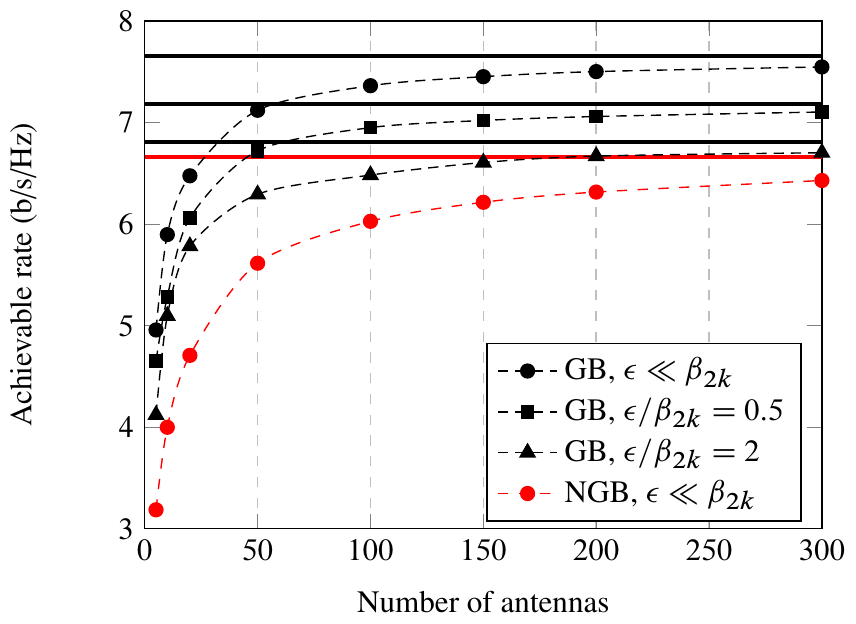}
\caption{Achievable rate (b/s/Hz) as a function of the number of antennas, in presence of non-negligible noise effects during the training phase. Scenario parameters: $L=2$, $K=1$, $\SNR=10$ dB, and $\beta_{11}/\beta_{21}=10$ dB (weak interference).\\}
\label{fig:figs3}
\end{figure}
\begin{figure}[!t]
\centering
\includegraphics{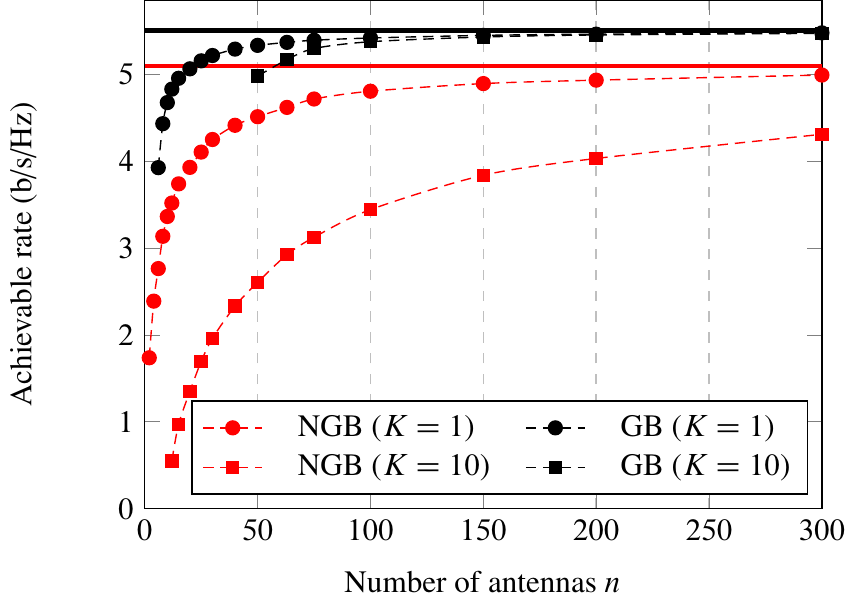}
\caption{Achievable rate (b/s/Hz) with group-blind (GB) and without group-blind (NGB) detection. Scenario parameters: $L=4$, $K=1$ or $K=10$, $\SNR=10$ dB, $\beta_{1k}/\beta_{2k}=10$ dB.}
\label{fig:fig7}
\end{figure}
%

In this section, we give numerical results to validate our analysis and show the performance gain achieved through group-blind detection. We assume $\beta_{11}=\cdots=\beta_{1K}=1$ and denote $\SNR=P\beta_{1k}=P$ the average received SNR from in-cell users. Solid lines on figures correspond to analytical results, whereas dashed lines connect simulation outputs. All figures confirm that simulations converge to our closed form expressions as the number of antennas grows. %

We consider a simple scenario in Fig.~\ref{fig:fig3}, with $L=2$, $K=1$, $\SNR=20$ dB, and $\beta_{11}/\beta_{21}=0$ dB (strong interference). We compare the achievable rate of non-group-blind (NGB) detection vs. group-blind (GB) detection. GB detection outperforms NGB detection by $\Delta\bar{R}_{1k}\approx 0.585$ (b/s/Hz). This value matches the asymptotic rate gap following from \eqref{eq:DeltaR}, \eqref{eq:newgammaapprox}, and \eqref{eq:sinrMF}, that is $\Delta\bar{R}_{1k}=\log_2(3)-\log_2(2)$ (b/s/Hz). 

Figure~\ref{fig:figs3} shows the achievable rate vs. the number of antennas $n$ for GB detection. We consider several values of $\epsilon/\beta_{2k}$, that model whether the estimation error is dominated by pilot contamination ($\epsilon<\beta_{2k}$) or thermal noise ($\epsilon>\beta_{2k}$). Figure~\ref{fig:figs3} is based on the following scenario: $L=2$, $K=1$, $\SNR=10$ dB, and fixed $\beta_{21}$ with $\beta_{11}/\beta_{21}=10$ dB (weak interference). The rate achieved with NGB detector in the presence of negligible noise during the training phase is plotted for comparison. The rate achieved with GB detectors decreases as $\epsilon$ grows, consistently with \eqref{eq:gammaL2}. However, even when the training phase is severely affected by noise, GB detection still outperforms NGB detection with noise-free training phase.


Finally, we consider in Fig.~\ref{fig:fig7} a scenario with $L=4$ cells, $\SNR=10$ dB, and a number of users per cell equal to either $K=1$ or $K=10$, and plot the achievable rate per user, as a function of the number of antennas $n$. GB detection is showed for $n\geq KL$, that is the required minimum number of antennas to implement the detector. In this case also, GB detection outperforms NGB detection. Moreover, the figure shows that GB detection is much more robust to variations of the network load, \ie, the number of users per BS antenna.

\section{Discussion}
We introduced a receiver for the uplink of multiuser massive MIMO that accounts for the presence of pilot contamination. The proposed scheme performs group-blind detection by exploiting the excess degrees of freedom provided by the large number of antennas per BS. We derived analytical results for the asymptotic achievable rate in an interference-limited scenario with one dominant interfering cell, and confirmed our findings through simulations. We found that group-blind detection outperforms traditional detection irrespective of the noise impairment during the training phase, and it is much more robust to variations of the network load. 

\vfill
\clearpage

\bibliographystyle{IEEEbib}
\bibliography{biblioguido_tran_pilots_ICASSP_16}

\begin{thebibliography}{10}

\bibitem{BocHeaLozMarPop:2014}
F.~Boccardi, R.W. Heath, A~Lozano, T.L. Marzetta, and P.~Popovski,
\newblock ``Five disruptive technology directions for {5G},''
\newblock {\em IEEE Commun. Mag.}, vol. 52, no. 2, pp. 74--80, 2014.

\bibitem{AndBuzWanetal:2014}
J.G. Andrews, S.~Buzzi, W.~Choi, S.V. Hanly, A.~Lozano, A.C.K. Soong, and J.C.
  Zhang,
\newblock ``What will {5G} be?,''
\newblock {\em IEEE J. Sel. Areas Commun.}, vol. 32, no. 6, pp. 1065--1082,
  2014.

\bibitem{Lueatl:2014}
L.~Lu, G.Y. Li, A.L. Swindlehurst, A.~Ashikhmin, and R.~Zhang,
\newblock ``An overview of massive {MIMO}: Benefits and challenges,''
\newblock {\em IEEE J. Sel. Topics Signal Process.}, vol. 8, no. 5, pp.
  742--758, 2014.

\bibitem{RusPerBuoLar:2013}
F.~Rusek, D.~Persson, B.K. Lau, E.G. Larsson, T.L. Marzetta, O.~Edfors, and
  F.~Tufvesson,
\newblock ``Scaling up {MIMO}: Opportunities and challenges with very large
  arrays,''
\newblock {\em IEEE Signal Process. Mag.}, vol. 30, no. 1, pp. 40--60, 2013.

\bibitem{HasHoc:2003}
B.~Hassibi and B.M. Hochwald,
\newblock ``How much training is needed in multiple-antenna wireless links?,''
\newblock {\em IEEE Trans. Inf. Theory}, vol. 49, no. 4, pp. 951--963, 2003.

\bibitem{JinLoz:2010}
N.~Jindal and A.~Lozano,
\newblock ``A unified treatment of optimum pilot overhead in multipath fading
  channels,''
\newblock {\em IEEE Trans. Commun.}, vol. 58, no. 10, pp. 2939--2948, 2010.

\bibitem{Mar:2010}
T.L. Marzetta,
\newblock ``Noncooperative cellular wireless with unlimited numbers of base
  station antennas,''
\newblock {\em IEEE Trans. Wireless Commun.}, vol. 9, no. 11, pp. 3590--3600,
  2010.

\bibitem{NgoLarMar:2013}
H.Q. Ngo, E.G. Larsson, and T.L. Marzetta,
\newblock ``The multicell multiuser {MIMO} uplink with very large antenna
  arrays and a finite-dimensional channel,''
\newblock {\em IEEE Trans. Commun.}, vol. 61, no. 6, pp. 2350--2361, 2013.

\bibitem{Josetal:2011}
J.~Jose, A.~Ashikhmin, T.L. Marzetta, and S.~Vishwanath,
\newblock ``Pilot contamination and precoding in multi-cell {TDD} systems,''
\newblock {\em IEEE Trans. Wireless Commun.}, vol. 10, no. 8, pp. 2640--2651,
  2011.

\bibitem{HoytenDeb:2013}
J.~Hoydis, S.~ten Brink, and M.~Debbah,
\newblock ``Massive {MIMO} in the {UL}/{DL} of cellular networks: How many
  antennas do we need?,''
\newblock {\em IEEE J. Sel. Areas Commun.}, vol. 31, no. 2, pp. 160--171, 2013.

\bibitem{GopJin:2011}
B.~Gopalakrishnan and N.~Jindal,
\newblock ``An analysis of pilot contamination on multi-user {MIMO} cellular
  systems with many antennas,''
\newblock in {\em Proc. IEEE Int. Workshop Signal Process. Adv. in Wireless
  Commun. (SPAWC)}, 2011, pp. 381--385.

\bibitem{FerAshMar:2013}
F.~Fernandes, A.~Ashikhmin, and T.L. Marzetta,
\newblock ``Inter-cell interference in noncooperative {TDD} large scale antenna
  systems,''
\newblock {\em IEEE J. Sel. Areas Commun.}, vol. 31, no. 2, pp. 192--201, 2013.

\bibitem{KriYatMan:2012}
N.~Krishnan, R.D. Yates, and N.B. Mandayam,
\newblock ``Cellular systems with many antennas: Large system analysis under
  pilot contamination,''
\newblock in {\em Proc. 50th Annu. Allerton Conf. Commun. Control Comput.},
  2012, pp. 1220--1224.

\bibitem{NgoLar:2012}
H.Q. Ngo and E.G. Larsson,
\newblock ``{EVD}-based channel estimation in multicell multiuser {MIMO}
  systems with very large antenna arrays,''
\newblock in {\em Proc. IEEE Int. Conf. Acoust. Speech Signal Process.
  (ICASSP)}, 2012, pp. 3249--3252.

\bibitem{MulCotVeh:2014}
R.R. Muller, L.~Cottatellucci, and M.~Vehkapera,
\newblock ``Blind pilot decontamination,''
\newblock {\em IEEE J. Sel. Topics Signal Process.}, vol. 8, no. 5, pp.
  773--786, 2014.

\bibitem{MulVehCot:2013}
R.R. Muller, M.~Vehkapera, and L.~Cottatellucci,
\newblock ``Analysis of blind pilot decontamination,''
\newblock in {\em Proc. Asilomar Conf. on Signals, Systems and Computers
  (ACSSC)}, 2013, pp. 1016--1020.

\bibitem{CotMulVeh:2013}
L.~Cottatellucci, R.R. Muller, and M.~Vehkapera,
\newblock ``Analysis of pilot decontamination based on power control,''
\newblock in {\em Proc. IEEE Veh. Technol. Conf. (VTC)}, 2013, pp. 1--5.

\bibitem{AndClaDohetal:2012}
J.G. Andrews, H.~Claussen, M.~Dohler, S.~Rangan, and M.C. Reed,
\newblock ``Femtocells: Past, present, and future,''
\newblock {\em IEEE J. Sel. Areas Commun.}, vol. 30, no. 3, pp. 497--508, 2012.

\bibitem{WanHos:1999}
X.~Wang and A.~Host-Madsen,
\newblock ``Group-blind multiuser detection for uplink {CDMA},''
\newblock {\em IEEE J. Sel. Areas Commun.}, vol. 17, no. 11, pp. 1971--1984,
  1999.

\bibitem{Kay:1993}
S.M. Kay,
\newblock {\em Fundamentals of Statistical Signal Processing: Estimation
  Theory},
\newblock Prentice-Hall, Inc., Upper Saddle River, NJ, USA, 1993.

\bibitem{Mad:1998}
U.~Madhow,
\newblock ``Blind adaptive interference suppression for direct-sequence
  {CDMA},''
\newblock {\em Proc. IEEE}, vol. 86, no. 10, pp. 2049--2069, 1998.

\bibitem{ZhaWan:2002}
J.~Zhang and X.~Wang,
\newblock ``Large-system performance analysis of blind and group-blind
  multiuser receivers,''
\newblock {\em IEEE Trans. Inf. Theory}, vol. 48, no. 9, pp. 2507--2523, 2002.

\bibitem{HosWan:2002}
A.~Host-Madsen and X.~Wang,
\newblock ``Performance of blind and group-blind multiuser detectors,''
\newblock {\em IEEE Trans. Inf. Theory}, vol. 48, no. 7, pp. 1849--1872, Jul
  2002.

\bibitem{XuWan:2004}
Z.~Xu and X.~Wang,
\newblock ``Large-sample performance of blind and group-blind multiuser
  detectors: a perturbation perspective,''
\newblock {\em IEEE Trans. Inf. Theory}, vol. 50, no. 10, pp. 2389--2401, 2004.

\end{thebibliography}

\end{document}